\documentstyle[aps,epsfig,prb,twocolumn]{revtex}
\begin{document}

\title{Theoretical study of the \protect\( (3\times 2)\protect \)
reconstruction of \protect\( \beta \protect \)-SiC(001)}

\author{L. Pizzagalli\cite{newadd}}

\address{Department of Physics and Astronomy, Basel University, 
Klingelbergstr. 82, CH-4056 BASEL,
Switzerland}

\author{A. Catellani}

\address{CNR-MASPEC, Parco Area delle Scienze, 37a, 43010 PARMA, Italy}

\author{G. Galli and F. Gygi}

\address{Lawrence Livermore National Laboratory, Livermore, CA 94551, USA}

\author{A. Baratoff}

\address{Department of Physics and Astronomy, Basel University, 
Klingelbergstr. 82, CH-4056 BASEL,
Switzerland
}
\date{received .....}
\maketitle

\draft

\begin{abstract}
 By means of \emph{ab initio} molecular dynamics and
band structure calculations, as well as using calculated  STM images,
we have singled out one structural model for the $(3\times 2)$
reconstruction of the Si-terminated (001) surface of cubic SiC,
amongst several proposed in the literature.
This is an alternate dimer-row model, with
an excess Si coverage of \(\frac{1}{3}\), yielding STM images in
good accord with recent measurements [F.Semond et al. Phys.~Rev.~Lett. 
{\bf 77}, 2013 (1996)].
\end{abstract}
\pacs{73.20.At,68.35.Bs}

\maketitle

The reconstructions of SiC(001) surfaces have been
widely studied in the last ten years\cite{Ber97NOT},
the characterization and understanding of growth mechanisms on the (001)
substrate being prerequisites for technological applications.
In the case of Si-terminated surfaces, several
reconstructions have been found to occur; \( (2\times 1) \), 
\( c(4\times 2) \)
and \( (n\times 2)_{n=3,5,7...} \) periodicities have been observed in
LEED\cite{Day85JVST,Kap89SS,Har90SSL},
RHEED\cite{Yos91APL} and 
STM\cite{Har94PRB,Sem96PRL,Sou97PRL2,Ari97PRL,Dou98SSL,Kit99SS}
measurements. Both (2\( \times  \)1) and c(4\( \times  \)2) reconstructions
pertain to a complete Si monolayer at the 
top \( (\theta _{Si}=1) \), as clearly
indicated by all available experimental data\cite{Har90SSL,Yos91APL}.
Unlike those of Si(001), the reconstructions of SiC(001) are 
characterized by
 weakly bonded, flat dimers ((2\( \times  \)1) \cite{powers,Cat98PRB})
or by alternating symmetric dimers with different heights
(c(4\( \times  \)2)\cite{Cat98PRB,Sou97PRL1,Piz98TSF}).
Adsorption of additional Si produces 
successive \( (n\times 2) \) reconstructions
as a function of Si coverage, including \( (7\times 2) \), \( (5\times 2) \),
and a combination of \( (5\times 2) \) and \( (3\times 2) \)
periodicities\cite{Har90SSL,Yos91APL,Har94PRB,Sou97PRL2,Dou98SSL,Kit99SS}.

The \( (3\times 2) \) reconstruction seems to be the last stage before
self-limitation of growth\cite{Har92SSa}. Its atomic configuration and
electronic structure are not clearly established, though they have been
intensively investigated. Three different atomic configurations, depicted on
Fig.~1, have been suggested in the literature. In the Double Dimer-Row (DDR)
model, proposed by Dayan\cite{Day85JVST} and possibly 
supported by other experimental
studies\cite{Har94PRB,Kit99SS,Yeo97PRB,Yeo98PRB}, there 
are two Si ad-dimers on top of
the full Si layer (Fig.~1.a). The resulting 
coverage \( \theta _{Si}=\frac{2}{3}\) is
in contradiction with the measured \( \theta _{Si} \) value 
of \( \frac{1}{3} \)
reported by several groups\cite{Har90SSL,Yos91APL,Har92SSa}.
The straightforward extension
of this model to the \( (5\times 2) \) reconstruction is also 
inconsistent with
the measured coverage\cite{Yos91APL}. Moreover, this model is not supported
by some STM studies\cite{Sem96PRL,Sou97PRL2}. The DDR is favored
by empirical molecular dynamics (MD) 
simulations\cite{Kit96APL}, but the Tersoff
potential used in these calculations is known to give a poor
description of \( \beta  \)-SiC surface 
reconstructions\cite{Tan95PRB}. Another
model, the ADDed dimer-row (ADD), was first suggested in an early study
by Hara et al\cite{Har90SSL}. This configuration, with one Si ad-dimer per
unit cell (Fig.~1.b), corresponds to the measured coverage for
the \( (3\times 2) \) and \( (5\times 2) \)
reconstructions. However, though it appears consistent 
with several experimental data,
both empirical\cite{Yan94SS} and \emph{ab initio}\cite{Yan95SS} calculations
have shown that it is not
energetically favored. Furthermore, STM investigations
do not support this model\cite{Har94PRB,Sem96PRL}. Another \( \frac{1}{3} \)
coverage model, the ALTernate dimer-row (ALT) (Fig.~1.c), was 
proposed by Yan et al\cite{Yan94SS,Yan95SS}.
This configuration is
supported both by calculations\cite{Yan94SS,Yan95SS} and by 
STM studies\cite{Sem96PRL,Sou97PRL2}.
However, it cannot account for the observed relation 
between single domain LEED
patterns with \( (2\times 1) \) and \( (3\times 2) \) 
periodicities\cite{Ber97NOT,Kap89SS}.
It also fails to explain the \( (3\times 1) \) reconstruction observed after
O or H adsorption\cite{Day85JVST,Har94PRB}. Note that all
three models involve Si ad-dimers which are perpendicular to the dimers on
the underlying Si surface.
Indeed, previous calculations have shown
that a single parallel ad-dimer
is energetically much less favored than a perpendicular
one\cite{Cat98APL}. 

All three models show some discrepancies either with existing experiments or
calculations,
but none of them can be safely ruled out, owing to the lack of consistency
between all available data.
In this contribution, we
report the results of self-consistent \emph{ab initio} total energy
calculations for all of the three structural models, including
full geometrical optimizations.  The surface energies
are compared using Grand Canonical Potentials. The
computed dispersion of electronic states, as well as STM
images, are compared to experiments.
Considering all the evidence, we conclude that
the \( (3\times 2) \) reconstruction of the Si-terminated surface
of SiC(001) is best described by the ALT model.

\begin{figure}
{\par\centering \resizebox*{6cm}{!}{\includegraphics{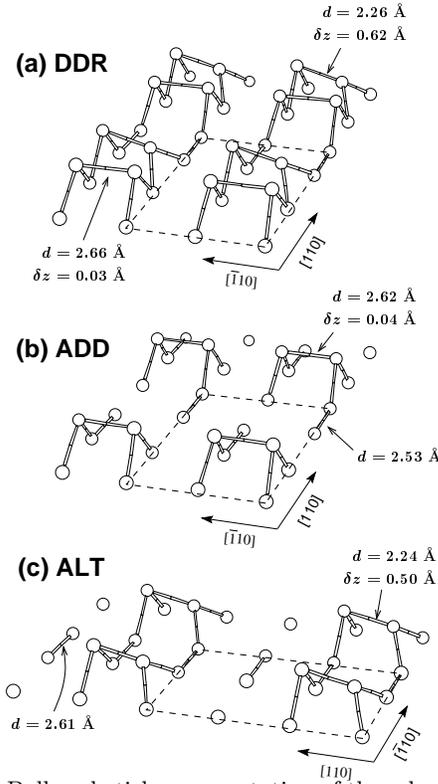}} \par}
\caption{Ball-and-stick representation of the relaxed atomic 
structures for the DDR
(a), ADD (b), and ALT (c) models. \protect\( d\protect \) 
and \protect\( \delta z\protect \)
are the distance and height difference between adatoms, 
respectively. Only the
ad-layer and the first underlying Si layer are shown for 
clarity. Bonds are drawn if the distance between atoms is smaller
than 2.7~\AA\ apart.}
\end{figure}

Our  calculations were performed at \( \text {T}=0\) within
the Local Density Approximation, using \emph{ab initio} molecular
dynamics codes employed in previous
studies of SiC surfaces\cite{Cat98PRB}.
Fully nonlocal norm-conserving pseudopotentials were used
for Si (s and p nonlocality) and C (s nonlocality)\cite{Ham89PRB}.
The system was simulated by a slab in a periodically repeated supercell. The
bottom layers were frozen in the p\( (2\times 1) \)
configuration determined earlier\cite{Cat98PRB}.
Two different sets of calculations have been performed. To determine
the relaxed surface atomic structures,
we used a \( (6\times 4) \) supercell with
8~atomic layers and a 10~\AA\ vacuum region (the total number
of atoms is 176 (184) for ALT and ADD (DDR) models).
The plane-wave energy cutoffs for the wave
functions and the charge density were 36~Ry and 130~Ry, 
respectively. Sums over occupied states
were performed at the \( \Gamma  \) point, which
corresponds to 4~inequivalent k-points in the Brillouin 
Zone (BZ) for a \( (3\times 2) \) cell. Next,
the electronic band structure was computed in a \( (3\times 2) \)
supercell with 12~atomic layers and a 6~\AA\ vacuum region, using an 
extension of ab-initio MD codes to finite wave functions vectors,
and keeping all atoms
fixed. Atomic positions in the six top layers and the two
bottom layers were taken from the preceding ab-initio MD calculations,
whereas those in the four central
layers were assumed bulk-like. Wave functions
and charge densities were expanded in plane waves with cutoffs of 40~Ry
and 160~Ry, respectively. In these calculations, the electronic 
charge density was computed using 8~special k-points in the
BZ, generated according to the Monkhorst-Pack scheme\cite{Mon76PRB}.

The relaxed atomic structures for the three models are shown on Fig.~1.
In the DDR geometry, one ad-dimer is strongly tilted (\(\delta z=0.62\)~\AA)
and has a short bond length (\(d=2.26\)~\AA) while the
other, weakly bound (\(d=2.66\)~\AA), is almost
flat (\(\delta z=0.03\)~\AA). The inequivalence of the 
two ad-dimers disagrees with
simple expectations\cite{Day85JVST,Har94PRB} and with previous 
calculations by
Kitabatake et al, who found two flat ad-dimers for the 
DDR model\cite{Kit96APL}.
Their use of the Tersoff potential could explain this disagreement,
owing to the neglect of charge transfer between Si and C atoms.
A single flat and weakly bonded ad-dimer (\(d=2.62\)~\AA)
is obtained in the ADD model, the geometry being close to that
previously obtained by Yan et al in a calculation similar 
to ours\cite{Yan95SS}.
Note that two slightly different configurations are possible 
within the same model,
since the weakly bonded Si dimers in the underlying layer
can be arranged either
all on one side, as originally proposed by Yan et al\cite{Yan94SS,Yan95SS},
or in a staggered
pattern (see Fig.~1). Starting from different 
configurations, our calculations
always converged to the staggered pattern\cite{Piz99NOT}. 
Finally, in the ALT model, the
ad-dimer is strongly tilted (\(\delta z=0.5\)~\AA)
and strongly bound  (\(d=2.24\)~\AA), in good
agreement with previous calculations\cite{Yan95SS}. The length of the weak
Si dimers in the underlying
surface layer is close to the value computed for 
the \( (2\times 1) \) reconstruction
using the same method and a \(4\times 4\) supercell\cite{Cat98PRB}.

In order to determine the most stable configuration, we have compared total
energies. The ALT model is lower in energy than the ADD model by
about 0.5~eV per \( (3\times 2) \) cell; this energy difference is clearly
in favor of the ALT, our error bar being 0.3~eV, as estimated from cutoff
and BZ sampling tests. The drastic reduction from the
3.6~eV energy difference quoted by Yan et al\cite{Yan95SS}
is likely due to their poorer BZ sampling and, to a lesser extent, to
the additional relaxation leading to the staggered pattern.
A direct comparison with the DDR model is not possible
since the corresponding Si coverage is different. This difficulty can be
overcome by using the grand canonical
scheme\cite{Qia88PRB}. The computed surface energy differences 
as a function of the
Si chemical potential \( \mu _{\text {Si}} \) are shown in
Fig.~2. The value of \( \mu _{\text {Si}} \)
for bulk silicon was calculated with an energy cutoff of 40~Ry 
and 32~special k-points
in the BZ, whereas we used the heat of formation of
SiC \( \Delta \text {H}=0.75 \)~eV from a recent 
calculation\cite{Nor95PRB}, to
obtain the Si chemical potential under
C-rich conditions. Since \( (3\times 2) \) growth is likely
to occur  under Si-rich conditions, a precise
determination of \( \mu _{\text {Si}} \) is required only in that limit.
Our results show that
the ALT model is the most stable configuration over the entire
allowed range of Si chemical potential. However, the energy 
difference between ALT and
DDR obtained under Si-rich conditions is only 77~meV, i.e. 
within our error bar.
Consequently, the DDR model can not be definitely ruled out 
solely on the basis
of total energy comparisons.

\begin{figure}
{\par\centering \resizebox*{6cm}{!}{\rotatebox{-90}{
\includegraphics{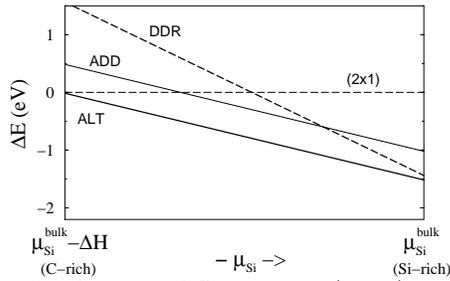}}} \par}

\caption{Total energy difference per \((3\times2)\) unit cell as a
function of the Si chemical potential \( \mu _{\text {Si}} \).}
\end{figure}

\begin{figure}
{\par\centering \resizebox*{6cm}{!}{\includegraphics{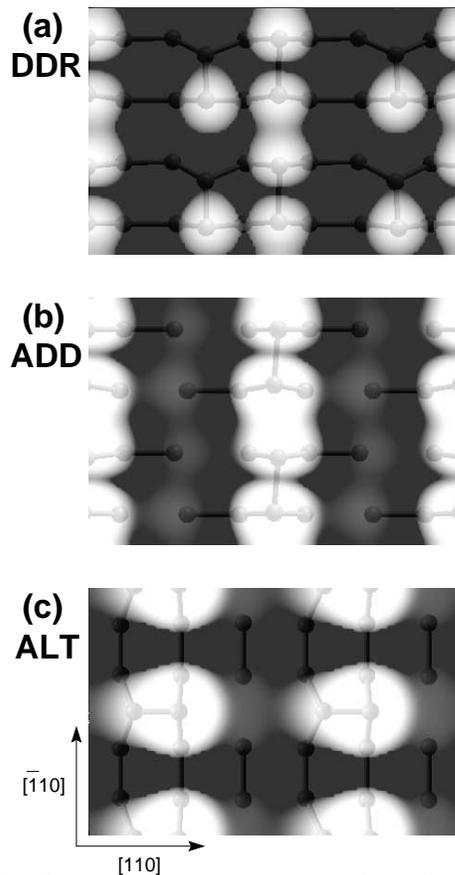}} \par}

\caption{Calculated constant-current STM images (bias \(V=-1\)\ V)
for the DDR (a), ADD (b), and ALT (c) models. Correlation with 
the atomic structure is explicitly shown via
the superposed ball-and-stick representations of the atoms in 
the first surface layer (see fig.~1).}
\end{figure}

Several experimental STM studies of the \( (3\times 2) \) 
reconstruction are currently
available\cite{Har94PRB,Sem96PRL,Har96SS,Har99SS}. In order to compare
the three different models, we have calculated
filled states constant-current STM images within the Tersoff-Hamann
approximation\cite{Ter85PRB}. Representative images are shown
in Fig.~3. In both the DDR and ADD models we find strings of
peanut-shaped spots, originating from a slight overlap between
maxima on adjacent flat ad-dimers. For the DDR model, additional
maxima are located on the up adatoms of the
tilted ad-dimers. The resulting images are incompatible 
with the experimental observations
of a single oval spot stretched in the \([110]\) direction per
\(3\times 2\) cell. On the other hand, in the ALT model
the spread out stretched spots located above up adatoms 
of the tilted ad-dimers
are in accord with experimental STM images of filled states\cite{Sem96PRL}.

Additional insight can be obtained from analysis of the electronic
states within a few eV of the Fermi level. All photoemission measurements
agree about the presence of two occupied surface states in the band gap,
1~eV apart from each
other\cite{Yeo98PRB,She94JVST,Lub98JVST}. However, uncertainties 
exist about the location of these states
with respect to the Valence Band Maximum (VBM). Recent 
angle-resolved photoemission
spectroscopy (ARPES) measurements have shown
that the dispersion of all identified surface states 
is very small (\(\leq 0.2\)~eV) along the
\( [\overline{{1}}10] \)\cite{Yeo98PRB} and 
\([110]\)\cite{Yeo98PRB,Lub98JVST} directions.
Only the surface states of the DDR and ALT models have been considered here,
the ADD model being higher in energy than the ALT model and
exhibiting STM images which do not agree with experiment.

We find that in the DDR model the surface is metallic,
within the Local Density Approximation.
The highest occupied state, about 1~eV above the VBM at \(\Gamma\),
is mainly localized on the flat ad-dimer
and has a \(\pi^*\) character with respect to the dimer axis. Its
dispersion is very small along the \([110]\) direction
(\(\leq 0.1\)~eV), but rather strong along
the \( [\overline{{1}}10] \) direction
(\(\simeq 1\)~eV).
In the DDR model, we also find three additional surface states
with energies between the highest occupied state and the VBM.
Only one of them is localized above the up adatom of the tilted
ad-dimer, and is essentially dispersionless.
The other two states originate from
backbond and dimer states of the underlying surface,
and show strong dispersions. The
presence of dispersive states
in the band gap is in disagreement with ARPES evidence
and points against the DDR model.

In the ALT model, the surface is semiconducting,
with a direct gap at \( \Gamma \)
of about 0.5~eV. The highest occupied state, 0.8~eV above the VBM
at \(\Gamma\), is localized on the
up adatom of the tilted ad-dimer and has a strong `s' character.
This surface state has a small dispersion along both
\([110]\) and \( [\overline{{1}}10] \)
directions (\(\leq 0.1\)~eV).
Close to the VBM we find another state, lying 0.7~eV below
the highest occupied orbital.
It is a \(\pi^*\) state localized on the Si-Si dimer of 
the underlying surface
which are not bonded to ad-dimers,
and is nearly dispersionless (\(\leq 0.2\)~eV). This
state is only present in the \([110]\) direction.
Except for the energy difference between the two highest
surface states (0.7~eV vs. 1~eV), agreement
with ARPES experiments is definitely better for the ALT than the DDR model.

A discrepancy exists
about the number and location of occupied surface states or
resonances\cite{Yeo98PRB,Lub98JVST}.
Yeom et al\cite{Yeo98PRB} argued that there should be a total
of four states with pronounced surface character
in their proposed DDR geometry. However, they considered the ideal
non-relaxed geometry,
with flat ad-dimers only, and, more importantly, they ignored
backbond states and dimer-like
states on the underlying surface which are also expected to exhibit surface
character. These additional resonant states
could be difficult to resolve because some are close in energy,
and might have weak
photoemission intensities. In our calculation we could identify resonant
states, however we did not attempt to systematically analyse
their character and dispersion.

Turning to unoccupied states, we find only one surface 
state in the band gap for the DDR model. It lies 2~eV
above the VBM at \(\Gamma\), is localized around the down adatom 
of the tilted ad-dimer and
has a predominant `p$_z$' character. Its dispersion is
about 0.1~eV (0.4~eV) along the \([110]\) (\( [\overline{{1}}10] \)) 
direction. For the ALT model, two
empty surface states have been identified. The lowest one, 
with energy 1.2~eV above the VBM at
\(\Gamma\), is a \(\sigma\)-like state on the lone Si dimer 
of the underlying surface. It disperses
about 0.1~eV along \([110]\) and 0.7~eV along \( [\overline{{1}}10] \). 
The other one, 1.8~eV
above the VBM, is a `p$_z$'-like state on the down adatom
of the tilted ad-dimer and has a very weak dispersion along
both directions (\(\leq 0.2\)~eV).

The combination of all our results indicates that the ALT model
is the most suitable
candidate, since it
explains the large majority of available measurements.
The ADD and DDR models produce incorrect STM images.
The ADD is energetically unfavorable, while the dispersion
of the surface states calculated for the DDR model is not 
compatible with ARPES measurements.

To summarize, we have performed plane-wave pseudopotential calculations for
three structural models of the
\((3\times2)\) reconstructed \(\beta\)-SiC(001) surface. In 
particular, relaxed atomic structures,
surface energies, STM images, surface states and their dispersion 
have been calculated and compared
with experiments. Our results strongly favor the ALT model and exclude
the DDR and ADD models, although some ambiguities remain. More definitive
conclusions could come
from additional experimental studies, in particular investigations of
unoccupied electronic states, and a convincing confirmation of the Si
coverage corresponding to the \((3\times2)\) reconstruction.

We are thankful to V.~M.~Bermudez, P.~Soukiassian,
G.~Dujardin and H.-W.~Yeom for fruitful
discussions and/or preprints. One of us (L.P.) gratefully
acknowledges Prof. H.-J. G{\"u}ntherodt for the facilitities
provided in his group,
and the Swiss National Foundation for financial support under the
NFP~36 program ``Nanosciences''.
This work has also been partially supported by
the "Consiglio Nazionale delle Ricerche" (Italy)  and
the Swiss Center for Scientific Computing (Manno, Switzerland).
Part of this work was performed by the Lawrence Livermore National Laboratory
under the auspices of the U.~S.~Department of Energy, Office of Basic
Energy Sciences, Division of Materials Science, Contract
No.~W--7405--ENG--48.

%\bibliographystyle{prsty}
%\bibliography{/home/pizza/NEW/articles/biblio/biblio,note}

\begin{thebibliography}{10}
\bibitem{Piz99NOT}
Staggered underlying dimers induce a slight rotation 
of the ad-dimers  towards the 
[110] direction. This resulting surface stress can be removed 
if one considers 
a staggered arrangement with mirror symmetry in the plane containing the 
ad-dimers, leading to a \((6 \times 2)\) reconstruction. This  geometry 
lowers the energy of the ADD model, but not enough to compete with the ALT 
model.
\end{thebibliography}

\end{document}